# Nonlinear Klein-Gordon equation fot nanoscale heat and mass transport

Janina Marciak-Kozłowska*, Mirosław Kozlowski ** ,

Magdalena Pelc

*Institute of Electron Technology Al Lotnikow 32/46,

02-668 Warsaw, Poland

Abstract

In this paper nonlinear Klein-Gordon equation for heat and mass transport in nanoscale was proposed and solved. It was shown that for ultra-short laser pulses nonlinear Klein-Gordon equation is reduced to nonlinear d`Alembert equation. The implicit solution of the d`Alembert equation for ultrashort laser pulses was obtained

**Corresponding author, e-mail :miroslawkozlowski@aster.pl

Introduction

The study of transport mechanisms at the nanoscale level is of great importance nowadays. Specifically, the nanoparticles and nanotubules have important physical applications for nano- and micro-scale technologies [1]. Many models have been developed in the simple picture of point-like particles  One possibility that has been considered in the literature is that of nonlinear Klein-Gordon system where the on-site potential is ratchet-like .

The development of the ultra-short laser pulses opens new possibilities in the study of the dynamics of the electrons in nanoscale systems: carbon nanotubes , nanoparticles. For attosecond laser pulses the duration of the pulse is shorter than the relaxation time. In that case the transport equations contain the second order partial derivative in time. The master equation is the Klein-Gordon equation

In this paper we consider the non - linear Klein Gordon equation for  mass and thermal energy transport in nanoscale. Considering the results of the monograph [1] we develop the nonlinear Klein Gordon equation for heat and mass transport  in nanoscale. For ultrashort laser pulse  the nonlinear Klein-Gordon equation is reduced to the nonlinear d`Alembert equation. In this paper we find out the implicit solution of the nonlinear d`Alembert equation for heat transport on nanoscale. It will be shown that for ultra-short laser pulses the non-linear Klein-Gordon equation has the nonlinear traveling wave solution

2. Nonlinear heat and mass transfer in  nanoscale

In monograph [1] it was shown that in the case of the ultra short laser pulses the heat transport is described by the Heaviside hyperbolic heat transport equation:

$$\tau \, \partial^2 T/\partial^2 t \; + \; \partial T/\partial t \; = \; D \, \partial^2 T/\partial^2 x \tag{1}$$

where T denotes the temperature of the electron gas in nanoparticle, $\tau$ is the relaxation time  m is the electron mass and D is the thermal diffusion coefficient. The relaxation time $\tau$ is defined as:

$$\tau = h/mv^2 \qquad v = \alpha c \qquad (2)$$

where $v$ is the thermal pulse propagation speed. For electromagnetic interaction when scatters are the relativistic electrons, $\tau$=Thomson relaxation time

$$\tau = h/mc^2 \qquad (3)$$

Both parameters $\tau$ and $v$ completely characterize the thermal energy transport on the atomic scale and can be named as "atomic" relaxation time and "atomic" heat velocity.

In the following, starting with the atomic $\tau$ and $v$ we describe thermal relaxation processes in nanoparticles which consist of $N$ light scatters. To that aim we use the Pauli-Heisenberg inequality [3]:

$$\Delta r \, \Delta p > N^{1/3} \, \hbar \qquad (4)$$

where $r$ denotes the radius of the nanoparticle and $p$ is the momentum of energy carriers.

According to formula (4) we recalculate the relaxation time $\tau$ for nanoparticle consisting N electrons:

$$\hbar^N \to N \, \hbar \qquad (5)$$
$$\tau^N = N \, \tau \qquad (6)$$

Formula (6) describes the scaling of the relaxation time for $N$ *fermion* systems.

With formulae (4) and (5) the heat transport equation takes the form:

$$\tau^N \partial^2 T/\partial^2 t + \partial T/\partial t = \hbar^{1/3}/m \, \partial^2 T/\partial^2 x \qquad (7)$$

and for mass transport:

$$\tau^N \partial^2 N/\partial^2 t + \partial N/\partial t = N^{1/3} \hbar/m \, \partial^2 N/\partial^2 x \qquad (8)$$

Equation (7) is linear damped Klein-Gordon equation, and was solved for nanotechnology systems in [1].

The nonlinearity of Eq. (8) opens new possibilities for the study of non-stationary stable processes in molecular nanostructures. Let us consider equation (8) in more details:

$$(N\tau)\partial^2 N/\partial^2 t + \partial N/\partial t = N^{1/3} \hbar/m \, \partial^2 N/\partial^2 x \qquad (9)$$

For $\Delta t < N\tau$ Eq (9) is the nonlinear d'Alembert equation

$$v^{-2} \partial^2 N/\partial^2 t = \partial^2 N/\partial^2 x \qquad (10)$$

with N dependent velocity:

$$v = N^{-1/3} \alpha c \qquad (11)$$

Equation (10) can be written in more general form:

$$\partial^2 N/\partial^2 t = \partial G(N)/\partial x \qquad (12)$$

where

$$G(N) = f(N) \, \partial N/\partial t \qquad (13)$$

The traveling wave solution of equation (12) has implicit form [5]:

$$\lambda^2 N(x,t) - \int G(N) dN = A(x + \lambda t) + B \qquad (14)$$

where A, B and $\lambda$ are arbitrary constants

References

[1]  M. Kozlowski, J. Marciak-Kozlowska, *Thermal processes using attosecond laser pulses,* Springer, 2006

[2] J.M. Levy-Leblond and F. Balibar, *Quantic,* North-Holland, 1990, p.445

[3] W F Ames et al., *Int. J. Nonlinear Mech.* vol.**16** (1981) 439